\documentstyle[mprocl]{article}

\def\Journal#1#2#3#4{{#1} {\bf #2}, #3 (#4)}

\def\PRD{{\em Phys. Rev.} D}
\def\MPL{{\em Mod. Phys. Lett.}A}
\def\GRG{\em Gen. Rel. Grav.}
\def\be{\begin{equation}}
\def\ee{\end{equation}}
\def\bea{\begin{eqnarray}}
\def\eea{\end{eqnarray}}

\begin{document}

\title{ QUANTUM COSMOLOGY IN SOME SCALAR-TENSOR THEORIES }

\author{ L.O. PIMENTEL\dag, C. MORA\dag\ddag }

\address{\dag Departamento de F\'{\i}sica, Universidad Aut\'onoma 
Metropolitana,\\
Apdo. Postal 55-534, CP 09340, M\'exico DF, M\'exico.}
\address{\ddag UPIBI-Instituto Polit\'ecnico Nacional, Av. Acueducto
s/n \\Col. Barrio La Laguna Ticom\'an, CP 07340 M\'exico DF, M\'exico.}

\maketitle\abstracts{
The Wheeler-DeWitt equation is solved for some scalar-tensor theories of
gravitation in the case of homogeneous and isotropic cosmological models.
We present general solutions corresponding to cosmological term: $(i)
\lambda(\phi)=0$ and $(ii) \lambda(\phi)=q\phi$. 
 }
 
\section {Introduction}

There is a renewed interest in the scalar-tensor theories of gravitation 
because of the unification theories and the chaotic inflation. Therefore it 
seems of interest to consider the quantum cosmology originated from these 
theories. We begin by introducing the action integral for the scalar-tensor 
theory with cosmological function $\lambda(\phi)$

\be
S=\frac{1}{16\pi}\int\!\sqrt{-g}\left\{\phi R -
\frac{\omega_0}{\phi} g^{\mu\nu}\phi_{,\mu}\phi_{,\nu} +
2\,\phi\,\lambda(\phi)\right\}\,d^4\!x, 
\label{eq1}
\ee

\noindent using the Friedmann-Robertson-Walker metric with positive
curvature, we can construct the corresponding Hamiltonian of the system.
After canonical quantization we obtain the Wheeler-DeWitt equation
$H\Psi(a,\phi)=0$ for an arbitrary factor ordering, encoded in the $\alpha$ 
and $\beta$ parameters,

\bea
\biggl\{\frac{\omega_0}{6}\left(a^2\frac{\partial^2}{\partial a^2} + \alpha
a\frac{\partial}{\partial a}\right) + a\phi
\frac{\partial^2}{\partial a\partial\phi}-\left(
\phi^2\frac{\partial^2}{\partial\phi^2}
+\beta\phi\frac{\partial}{\partial\phi}\right)
{}-\nonumber\\ 
 -\frac{\pi^2}{16}(2\omega_0 + 3)\left[3a^4\phi^2 -
 a^6\phi^2\lambda(\phi)\right]\bigg\}\Psi(a,\phi)=0.
\label{eq2}
\eea

\section{Solutions for the Wheeler-DeWitt equation with $\lambda(\phi)=0$ 
and $\lambda(\phi)=q\phi$}

We solve the WDW equation (\ref{eq2}) by separation of variables. The
corresponding solutions are show in the following cases

\noindent {\it i)} $k^2>B$ 

\be
\Psi_k(a,\phi)=\phi^{-\frac{A}{2}}
a^{A}\sum_{m=\mp 1 \atop n=1,2}\phi^{mi
\sqrt{k^2\rho^2-\left(\frac{\beta-1}{2}\right)^2}}
C_nH_p^{(n)}\left(\frac{3\pi}{4}ia^2\phi\right)
\label{eq3}
\ee
 
\noindent {\it ii)} $k^2=B$ 

\be
\Psi_k(a,\phi)=\phi^{-\frac{A}{2}}
a^{A}(C_3+C_4\rho\ln\phi) \sum_{n=1,2}
C_nH^{(n)}_p\left(\frac{3\pi}{4}ia^2\phi\right),
\label{eq4}
\ee

\noindent {\it ii)} $k^2<B$

\be
\Psi_k(a,\phi)=\phi^{-\frac{A}{2}}
a^{A}\sum_{m=\mp 1 \atop n=1,2}\phi^{mi
\sqrt{\left(\frac{\beta-1}{2}\right)^2-k^2\rho^2}}
C_nH_p^{(n)}\left(\frac{3\pi}{4}ia^2\phi\right)
\label{eq5}
\ee

\noindent $k$ is a separation constant, $H^{(1,2)}_p$ are the Hankel 
functions of order $p$, $B=\frac{\beta-1}{2\rho}$
$A=\frac{\omega_0(\alpha-1)-3(\beta-1)}{2\omega_0+3}$,
$\quad\rho^2=\frac{2\omega_0+3}{3},
\quad\hbox{and}\quad p=\sqrt{\frac{A^2}{4}-k^2}.$

\noindent If we consider an specific factor ordering $\alpha=\beta=1$ in 
eq.(\ref{eq3}), and making superposition of wave functions we have

\be
\psi_k=e^{-\frac{3\pi}{4}a^2\phi\cosh[\rho\ln\phi+\mu]}.
\label{eq6}
\ee

\noindent This solution satisfied the Hawking-Page regularity 
condition~\cite{Haw}, {\it i.e.}, {\it 1)} the wave function is 
exponentially damped for large spatial geometry $(a \to \infty)$, 
and {\it 2)} the wave function is regular when the spatial geometry 
degenerates ($\Psi(a,\phi)$ does not oscillate when $a \to 0$), 
thus eq. (\ref{eq6}) can be regarded like quantum wormhole 
solutions~\cite{Cav}.

\noindent Now, we present solutions of WDW eq. when $\lambda(\phi)=q\phi$, 
  where $q$ is a constant, this equation is solvable by series, but it 
  solution haven't physical meaning. There is an interesting  exact solution 
  case with the next condition

\be
\omega^2_0(\alpha-1)^2-6\omega_0(\alpha-1)(\beta-1)+9(\beta-1)^2=
(1-4k^2)(2\omega_0+3)^2,
\label{eq7}
\ee

\noindent {\it i)} $k^2>B$ 

\be
\Psi_k(a,\phi)=\sum_{{m=\mp 1 \atop n=1,2} \atop L=Ai,Bi}
\phi^{\frac{\beta-1}{2}+mi\sqrt{k^2\rho^2 - 
\left(\frac{\beta-1}{2}\right)^2}}
C_nL\left[\frac{4}{\pi\sqrt{3q^3}}\left(3-qa^2\phi\right)\right],
\label{eq8}
\ee

\noindent {\it ii)} $k^2=B$ 

\be
\Psi_k(a,\phi)=\phi^{\frac{\beta-1}{2}}
(C_3+C_4\rho\ln\phi)
\sum_{m=1,2 \atop L=Ai,Bi}
C_mL\left[\frac{4}{\pi\sqrt{3q^3}}
\left(3-qa^2\phi\right)\right],
\label{eq9}
\ee

\noindent {\it iii)} $k^2<B$ 

\be
\Psi_k(a,\phi)=\sum_{{m=\mp 1 \atop n=1,2} \atop L=Ai,Bi}
\phi^{-\frac{A}{2}+mi
\sqrt{\left(\frac{\beta-1}{2}\right)^{2}-k^2\rho^2}}
C_nL\left[\frac{4}{\pi\sqrt{3q^3}}
\left(3-qa^2\phi\right)\right],
\label{eq10}
\ee

\noindent where $Ai$ and $Bi$ are the Airy functions.

\section{Final remarks}
We have found solutions of WDW equation in BD theory, for two cosmological
functions:  $\lambda(\phi)=0$ and $\lambda(\phi)=q\phi$, and we want to
note that there are quantum WH solutions in the first case, i.e., when
cosmological constant vanish, this agree with results obtained by 
Xiang {\it et al.}~\cite{Xia} using a particular factor ordering.

\section*{Acknowledgments}

This work has been supported by CONACYT and I.P.N.

\end{document}